\documentclass[showpacs,twocolumn,eqsecnum,aps, amssymb,nofootinbib,prd]{revtex4}
\usepackage{graphicx, epsfig, times, bm, mathptm}

\newcommand{\be}{\begin{equation}}
\newcommand{\ee}{\end{equation}}
\newcommand{\bea}{\begin{eqnarray}}
\newcommand{\eea}{\end{eqnarray}}

\newcommand{\PR}{{\it Phys. Rev.\,}}

\newcommand{\PL}{{\it Phys. Lett.\,}}

\begin{document}

\title{Cosmological Implications of the String Theory Landscape}

\author{L.~Mersini-Houghton}

\affiliation{Department of Physics and Astrononmy, UNC-Chapel Hill, NC, 27599-3255, USA}

\email[]{mersini@physics.unc.edu}

\date{\today}
 
\begin{abstract} 
Progress in string theory has resulted in a whole landscape of vacua solutions.In this talk I describe a proposal for exploring the cosmological implications of the landscape, based on the dynamics of the wavefunction of the universe propagating on it. 
The landscape is taken as the phase space of the initial conditions since every of its vacua can potentially give rise to a universe. A superselection rule on the landscape phase space emerges when we include the backreaction of massive long wavelengths  and the quantum dynamics of gravitational and matter degrees of freedom. The quantum dynamics of the system selects only high energy patches which survive the collapse induced by the gravitational instability of massive perturbations.

{\it Talk given in the workshop 'The Dark Side of the Universe', 2006, Madrid}

\end{abstract}

\pacs{98.80.Qc, 11.25.Wx}

\maketitle

\section{Introduction}


Recent advances in string theory and cosmology have placed fundamental questions about our universe in the forefront of research. Besides the outstanding puzzle of the nature and origin of dark energy, the selection of the initial conditions for our universe remains one of the deepest mysteries in nature. Every form of energy gravitates and each species couples to the others gravitationally.Self-gravitating properties of dark matter and dark energy provide us with evidence for their existence. In the light of the problems we are facing, a deeper understanding of the quantum dynamics of gravity becomes a prerequisite. It is wildly accepted that some sort of a larger scale dark energy $\Lambda_i$ was also responsible for even starting our universe, refered to as the Big Bang inflation. The inflationary theory seems so far in perfect agreement with observational data. However it neither addresses why our universe started with such initial conditions of low entropy, nor how the physics of gravity, the quantum and thermodynamics tie in together in this picture. By this, I specifically mean that Big Bang inflation does not address the friction between some of the underlying principles of the above theories, namely: causality, unitarity and the second law of thermodynamics. Therefore it is not as yet a complete theory.

Simple statistical arguments tell us that since the entropy of the inflating patch is $S_i\simeq \frac{3}{\Lambda_i}$ and the probability for starting with these initial conditions is $P\simeq e^{S_i}$ then, by the second law of thermodynamics, either our initial conditions are extremely unlikely, therefore special, or we do not understand the arrow of time determined by the direction of the entropy growth. Clearly, the selection of the initial conditions which seems at odds with the arrow of time, is one of the most fundamental problems in physics.
Meanwhile a unitary evolution requires that if we started with an initial mixed state then we can not evolve into a pure state\cite{shenker,richlaura1,richlaura2}. Causality would treat our patch as an isolated system. When the principle of causality, seemingly at odds with unitarity \cite{susskind,shenker}, is applied to a $\Lambda$CDM universe such as our universe, it introduces a bound on the high energy initial state $\Lambda_i$ in terms of the low energy horizon scale of the present universe, $H_0\simeq \sqrt{\Lambda_0}$\cite{susskind2,entropy1}. There is little hope that we would make sense of our initial and present universe or understand where these puzzling paradoxes stem from, without a deeper understanding of gravity itself. The vacuum energy $\Lambda$ may be fundamentally quantum in nature.Hence its self-gravitating properties may be weaving the fabric of spacetime itself, a self-evident feature were we to move $\Lambda$ to the left hand side of Einstein equations that describe the geometry.A theory of quantum gravity is needed for a deeper understanding of these fundamental issues. At present, string theory is a serious contender for quantum gravity. For this reason, exploring the cosmological implications of string theory is well worth the effort. Furthermore, the cosmological arena is expected to provide the 'playground' for testing predictions of string theory, (e.g.\cite{katie,steenlaura}).

Major efforts in recent years, in connecting string theory to the observable universe, have resulted in a landscape picture with a multitude of solutions.Basically starting from higher dimensions, the vacua solutions found after dimensional reductions that contain $(3+1)-D$ worlds like ours, are not unique but there are more like $10^{500}$ of them. The emerging landscape of string vacua has been considered by many as bad news for both string theory and cosmology. From the string theory perspective, the landscape appeared to imply that the theory may lose its predictive powers by producing enough solutions as to become unfalsifable. From a cosmological perspective the implication seemed to be that, in this vastness, the selection of the vacua which hosts our universe may be an impossible task. 

To my opinion, the emerging landscape picture is not bad news at all for string theory and cosmology.{\it Au contraire} I consider it to be a very good indicator that string theory may be a serious candidate for quantum gravity, on the basis that a landscape picture {\it must} be expected of any contender for the quantum gravity throne as we argued in \cite{laurareview,land1,land2,flysoup,richlaura1,richlaura2}. Let me summarize the argument here: Any theory of quantum gravity must contain a landscape, a metauniverse, of initial patches in order to provide us with a physical phase space for the initial conditions from which we can select ours. But since every site on the landscape is a potential starting point for giving birth to a universe then the ensemble of these solutions that make the landscape thus becomes the phase space for the initial conditions. A theory of quantum gravity should not yield just a {\it unique sample} because fundamental questions about our universe do require the existence of a phase space, a.k.a the landscape. I do not see how else we could meaningfully ask questions about the choice of the initial conditions, constants of nature, etc. in our universe, from within one sample, i.e. without implying: as compared to what other choices\footnote{I would like to thank Ch.Isham for very helpful discussions on this point}? A theory of quantum gravity is thus expected to provide us with an ensemble of possible initial conditions and quantum numbers.
Based on this point of view, we argued \cite{richlaura1,laurareview,land2} that a landscape will emerge out of any theory of initial conditions, the theory of quantum gravity. 

However, deriving a phase space for the initial conditions, from the underlying theory, is the first stage expected of the theory. Within this landscape, we still need to address the second issue, namely what {\it criterion} selected the initial conditions for our patch in this vastness. The hope is that the quantum dynamics of these initial patches, with gravitational and matter degrees of freedom,produces a {\it superselection rule} whereby the Universe finds itself driven to choose a {\em unique} vacuum state. We proposed in \cite{land1,land2,laurareview} that such a selection may be computed by the requirement that this solution be the most probable state the Universe can access \cite{land1,land2,laurareview,richlaura1,richlaura2}.

This line of reasoning led us to propose the approach of allowing the wavefunction of the universe propagate on the landscape of string theory, with the latter as a candidate for the quantum theory of gravity \cite{land1,land2,laurareview,richlaura1,richlaura2}. The expectation in this program is that investigating the quantum dynamics of gravity would shed light into the deep puzzling problems such as how did our universe start and question which one of our assumptions about the early universe may not be warranted. In this setting, string theory would still be as predictive as other theories in physics, provided it required a much deeper understanding of the initial conditions (IC) for the Universe. A dynamical selection of the initial state results in a reduction of the number of the allowed initial states\cite{richlaura1}. In this talk I describe how this superselection criterion emerges from decoherence obtained by the backreaction of matter modes onto the gravitational degrees of freedom.
 
The dynamical mechanism exhibited below is the following: the inclusion of the backreaction due to massive scalar perturbations gives rise to instabilities that render most of the inflationary patches {\em unstable} against gravitational collapse of super-horizon modes. This has the effect of dynamically reducing the allowed phase space of stable inflationary patches. The mechanism is essentially a Jeans instability effect, arising from the generation of tachyonic modes by the backreaction of the perturbations in Wheeler-deWitt (WdW) Master equation. We can then trace out the modes corresponding to collapsing patches to construct a reduced density matrix $\rho_{\rm red}$ for the patches that survive the collapse and enter an inflationary phase. We use this to show explicitly that if $\hat{H}$ is the Hamiltonian of the system then $\rho_{red}$ evolves with time, $\left[\hat{H},\rho_{\rm red}\right]\neq 0$. An important implication of our results is that the phase space of the initial states allowing for inflation is {\em not} ergodic since its original volume is reduced to only a subset that contains survivor universe only. Most of the details can be found in \cite{richlaura1,richlaura2,land1,land2,laurareview}. 


\section{Wavepackets on the Landscape and Wheeler-DeWitt Equation}
\label{sec:land}

Given that the structure of the landscape is as yet largely unknown, we capture the features of the landscape that might be important for discussing inflationary initial conditions by using the vacua distribution given in \cite{douglas}.This probability distribution of vacua is analogous with the CI-type Universality class of condensed matter systems \cite{altland}, (quantum dots), that have rotational and time invariance symmetries. This model incorporates both, the moduli internal degrees of freedom in each vacua, (resonances closely spaced around the mean value of the moduli label for that vacuum state) as well as the collective coordinate $\phi$ which labels the vacua sites from one another.

We will follow the minisuperspace approach, where besides the landscape degree of freedom $\phi$, we need to include the gravitational degree of freedom for each universe, represented by the scale factor $a$.  Time invariance is now broken and the non-SUSY sector of the landscape, with gravity switched on, belongs in the type C universality class. This allows us to write the joint probability distribution of vacua as \cite{altland}:

\begin{equation}
P( \langle \hat{H}(\phi) \rangle)= P(\omega^2)\approx M_{\rm P}^{-8} \prod_{i\le j}({\omega_{i}^2}
 - {\omega_{j}^2})^2 \prod_k \omega_k^{2}\ e^{-\frac{\omega_k^2}{v^2}}.
\label{jointprobab}
\end{equation}
where $\epsilon_i = \omega_i^{2}$.
In the limit that the energy level spacing is less than $b=v\sqrt{M}$, where $M$ is the number of the internal degrees of freedom/sublevels in the $i$'th vacuum, (the closely spaced string resonances around the $i$'th vacua), this result goes to the familiar Wigner-Dyson result of random disordered systems, $P(\langle\hat{H}(\phi)\rangle =\omega^2) \approx \omega^2$.  We also see that for large energies, $P\approx (\omega^2 +\gamma - v^2)e^{-\omega^2(1/v^2 + l)}$.


We now allow the wavefunction of the universe \cite{land1,land2} to propagate in this structure. The landscape is treated as a {\em disordered} lattice of vacua, where each of the $N$ sites is labelled by a mean value $\phi_i,\ i=1,\dots N$ of moduli fields. We make use of the Random Matrix Theory (RMT) \cite{review1,efemetov} for finding solutions to the Wheeler DeWitt (WDW) equation. The disordering of the lattice is enforced via a stochastic distribution of mean ground state energy density $\epsilon_i,\ i=1\dots N$ of each site. These energies are drawn from the interval $\left[-W,+W\right]$, where $W\sim M_{\rm Planck}^4$ with a Gaussian distribution with width (disorder strength) $\Gamma$: $M_{\rm SUSY}^8\lesssim \Gamma\lesssim M_{\rm Planck}^8$, where $M_{\rm SUSY}$ is the SUSY breaking scale. Quantum tunneling to other sites allows the wavefunction to spread from site to site. The stochastic distribution of sites ensures Anderson localization\cite{anderson} of wavepackets around some vacuum site, at least for all the energy levels up to the disorder strength.  The energy density of the Anderson localized wavepacket is $\epsilon_i =|\Lambda_i +i\gamma|$, where $\Lambda_i$ is the vacuum energy density contribution to the site energy $\epsilon_i$ and $\gamma=l^{-1}$, where $l$ is the mean localization length and $l_p$ is the fundamental length of the lattice, which we will be take to be the Planck/string length. For large enough values of the disorder strength $\Gamma$, the majority of the levels are localized so that a semiclassical treatment of their classical trajectories in configuration space is justifed. 
The single-particle density of states $\rho(\omega)=\langle {\rm Tr} \delta(\omega^2 -H(\phi))\rangle$, obtained by integrating the above joint probability with respect to $\omega$, behaves as $\rho(\omega)\propto M_{\rm P}^{-8}  (1-\sin(l \omega^2)/l\omega^2)\ e^{-\frac{\omega^2}{v^2}}$. When time reversal symmetry, given by the operation $\epsilon \rightarrow -\epsilon$, is broken, then $\rho(\omega)\simeq (1+\sin(l \omega^2)/l\omega^2)\ e^{-\frac{\omega^2}{v^2}}$.These results indicate that the most probable solutions are peaked around zero energy vacua, (see Fig.[1,2] in Ref.\cite{richlaura1} for plots of the above density of states).

The landscape minisuperspace is spanned by $a,\phi$ the scale factor of 3 geometries and the collective coordinate of the landscape respectively,\cite{DeWitt}.The wavefunction $\Psi$ depends on the scale factor $a(t)$, curvature $\kappa=0,\ \pm1$ of the FRW 3-geometries together with the landscape variables, collectively denoted by $\left\{\phi\right\}$ which will play the role of the massive modes in the Wheeler-DeWitt (WDW) equation. We allow $\Psi$ to propagate on the landscape background with the vacuum distribution described above and parametrized by the collective coordinate $\{\phi\}= \{\phi_i^{n}\}$ where $\phi_i$ is the central value of landscape variable on vacuum site $i=0,1,2,...N$, and $n$ counts the internal degrees of freedom within the $i$'th vacuum. We can think of $n$ as counting the sublevels within the $i$'th energy level with width $b= vM $, and of the $\phi_1,...\phi_N$ as distinct energy levels.

The Wheeler-DeWitt equation for the wavefunction of the universe propagating on the minisuperspace spanned by the landscape variable $\phi$ and the FRW 3-geometries with line element
\begin{equation}
ds^2= -N^2 dt^2 + a^2(t) d{\bf x}^2 ,
\label{eq:lineelt}
\end{equation}  
is \cite{qcreview,wdw} 

\begin{eqnarray}
& &{\hat {\cal H}}\Psi(a,\phi) = 0 ~{\rm with} \nonumber \\
& &\hat{{\cal H}}=\frac{1}{2e^{3\alpha}}\left[\frac{4\pi}{3M_p^2}
\frac{\partial^2}{\partial\alpha^2}-
\frac{\partial^2}{\partial\phi^2}+e^{6\alpha}V(\phi)\right].
\label{eq:wdweq}
\end{eqnarray}
Here the scale factor $a$ has been written as $a=e^{\alpha}$ and $V(\alpha, \phi) = e^{6\alpha}V(\phi) - e^{4\alpha}\kappa, \kappa =0, 1$ for flat or closed universes.

The wavefunction $\Psi(\alpha,\phi)$ will in general be a superposition of many waves. In order to build wavepackets that correspond to classical paths in configuration space, some form of decoherence has to occur. Usually, this requires a separation between ``system'' and ``environmental'' variables; tracing over the environmental variables converts the ``system'' into an open one and allows it to behave classically.
Let us first turn to the construction of wavepackets centered around a vacuum characterized by $\phi_i$~\cite{kiefer}. Using this wavepacket, we will then include the backreaction of the environment modes on this wavepacket. This will lead us from the WdW equation to a {\it Master Equation} for the wavefunction $\Psi(\alpha,\phi)$.
 
When we specialize these results to our version of the landscape, we consider the rescaled variables $ x=e^{3\alpha}\phi$, $\omega_k^{2}=e^{6\alpha}\ \omega_k^{2}$ which leads to
\begin{eqnarray}
\hat{\cal{H}}(x)\psi_{j}(x) = \hat{\epsilon}_j \psi_{j}(x) ~{\rm where} \nonumber  \\
\hat{\cal{H}}(x)=\frac{3M_p^2}{4\pi} \left[
\frac{\partial^2}{\partial x^2} - (\omega_k^{2}-\gamma)\right] \nonumber \\
\frac{\partial^2}{\partial \alpha ^2}F_j(\alpha) + (\omega_j^{2} - \gamma + \kappa e^{4\alpha})F_j&=&0 .
\label{eq:landwdweqn}
\end{eqnarray}

The localized solutions $\psi_j(x)$ around a vacuum site with energies centered around $\tilde{\omega_j}$ within the gaussian width $v$, are
 
 \be
 \psi_j (x)\simeq \sin(\omega_j x) \ e^{-\frac{(x-x')}{l} }.
 \ee 
 
The wavepacket is constructed as a superposition of these solutions for the $M$ internal degrees of freedom $n=1,...M$ with energies peaked around the mean value of site $x_j$, $\epsilon_j$, and amplitudes given by the Gaussian weight 
$$
A^{j}_{n} = \frac{l}{\pi\sqrt{Mv^2}}\ e^{- (\omega_n-\omega_0)^2 /(Mv^2)},
$$
namely\footnote{We will drop the index $j$ that labels the site from now on, keeping only the index $n$ that counts the internal degrees of freedom of the $j$'th  site}, 

$$
\psi(\omega_j)=\int_n d {\omega_n} A_n \psi_n F_n(\alpha).
$$
Tracing out the internal perturbation modes described by the index $M$ results in the reduced density matrix \cite{kiefer} for the system $(\alpha,x)$. The first term $\rho_0(a,a')$ in Eq.~\ref{eq:density2} shows that the intrinsic time $a$ of the wavepacket  becomes classical first since the internal number of degrees of freedom $M$ is large, while $\phi$ becomes classical later when the scale factor grows larger than the Gaussian width below
\begin{equation}
\rho\sim e^{-\frac{\Omega_{c1} M}{1}(a-a^{\prime})^2} \\
e^{-(b^2 - \frac{b^4}{4\gamma^2})^2 a^6 (\phi-\phi^{\prime})^2}.
\label{eq:density2}
\end{equation}
with $\Omega_{cl} = (m_{0}/M)^{1/2}$.
The reduced density matrix above indicates how well the mean value $\epsilon_i,\phi_i$ can describe the vacuum site $i$ when the energy levels broaden due to the internal fluctuation modes of $\phi_i$. We expect the width $b=v\sqrt{M}$ to be at least of order SUSY breaking scale $M_{\rm susy}$, in order to account for the SUSY breaking of the zero energy levels $\omega_k=0$. Since the Fourier transform of the above wavepacket is still a Gaussian with width inverse that of $(x-x_0)^2$, we need  $b^2 < 2\gamma $ or $M_{\rm susy}\le M_*$, in order to have a meaningfully centered energy for the wavepacket made up from all the internal degrees of freedom $i=1,2,..M$. However, this gives rise to a spreading of the wavepacket in the moduli space $x$. To classicalize the system, we need to include the higher multipoles as environmental variables. 
These results show that, within the WDW formalism on a $2$ dimensional minisuperspace approximation, the most probable universe seems to be selecting the zero energy vacua on the landscape, if we do not include the effect of backreaction. What happened to high energy inflation and how did our patch decohere from the others? Let us discuss this below by including the backreaction term in the WDW equation.

\section{Master Equation and Decoherence}
\label{subsec:backreact}
 It is clear that the quantum mechanical probability measure, $P\simeq |\Psi|^2$, is not sufficient for addressing the selection of high energy initial states. We need to define an observer who measures and decoheres our patch from the others. At this stage we have implicitly assumed that if string theory is the theory of quantum gravity beyond general relativity, then the string landscape becomes the arena or the 'metauniverse' for the ensemble of the initial conditions. Proposing that the landscape is the phase space of the initial conditions allows us to select a superobserver that  'watches' the landscape multiverse,\cite{laurareview}, and can ask fundamental questions such as:Why did our universe pick this particular initial condition, quantum numbers and constants of nature, in such vastness of possibilites in the phase space. 

The moduli fields as well as the metric have fluctuations about their mean value and those fluctuations can serve to decohere the wavefunction\cite{halliwellhawking}. This would then provide a classical probability distribution for scale factors and moduli fields. The procedure laid out in Ref.\cite{halliwellhawking} starts by writing the metric and the moduli fields as
\be
\label{eq:pert}
h_{ij} = a^2 \left(\Omega_{ij}+\epsilon_{ij}\right),\ \phi=\phi_0+\sum_n f_n(a) Q_n,
\ee
where $\Omega_{ij}$ is the FRW spatial metric, $\epsilon_{ij}$ is the metric perturbation (both scalar and tensor), $Q_n$ are the scalar field harmonics in the unperturbed metric and $f_n(a)$ are the massive mode perturbations.  For our model of the landscape, we will take the higher super-horizon wavelength massive and metric multipoles $\{f_n,d_n\}$ to play the role of environmental variables. These modes couple with gravitational strength to the system $\Psi(\alpha,\phi)$. This coupling is of order $g\simeq{GM}\slash{R}$ with $M\approx O(M_{\rm Jeans}) \simeq H$ and $R\approx r_H\simeq H^{-1} $ so that $g\simeq H^2\slash M_P^2$. This is usually very small so that we can treat the higher multipoles as environmental variables and trace them out perturbatively~\cite{halliwellhawking,kiefer}.  
The index $n$ is an integer for closed spatial sections, and $k=n\slash a=n e^{-\alpha}$ denotes the physical wavenumber of the mode. As shown in Ref.\cite{halliwellhawking}, the fact that the CMB fluctuations are so small means that we can neglect the effects of the metric perturbations $\{d_n\}$in the following calculations, relative to the field fluctuations $\{f_n\}$.

The wavefunction $\Psi=\Psi(a, \phi, \left\{f_n\right\})$ is now a function on an infinite dimension minisuperspace spanned by the variables $\{a,\phi,f_n\}$. Inserting Eq.(\ref{eq:pert})   into the action, yields Hamiltonians $\left\{H_n\right\}$ for the fluctuation modes, which at quadratic order in the action, are decoupled from one another. The full quantized Hamiltonian $\hat{H} = \hat{H}_0 + \sum_n {\hat{H}}_n$ then acts on the wavefunction
\be
\label{eq:wavepert}
\Psi \sim \Psi_0 (a, \phi_0) \prod_n \psi_n (a, \phi, f_n).
\ee 
Doing all this yields the master equation
\be
\label{eq:master}
\hat{H}_0 \Psi_0 (a, \phi_0) = \left(-\sum_n \langle  \hat{H}_n\rangle\right) \Psi_0 (a, \phi_0),
\ee
where the angular brackets denote expectation values in the wavefunction $\psi_n$ and
\be
\label{eq:nham}
\hat H_n = -\frac{\partial^2}{\partial f_n^2} + e^{6 \alpha} \left( m^2 +e^{-2 \alpha} \left(n^2-1\right)\right) f_n^2,
\ee

The procedure laid out here also provides us with a 'guestimate' of the quantum entaglement of the initial state (see \cite{richlaura1,richlaura2} for details).This in turn enables us to check the validity of any assumptions made about ergodicity of the phase space, when the dynamics of the gravitational degrees of freedom is taken into consideration and probe deeper into the friction between causality,unitarity and thermodynamics \cite{richlaura2,entropylaura}.

\subsection{Superselection in the Phase Space of Inflationary Patches}
\label{sec:select}
Following Ref.\cite{kiefer,halliwellhawking} a time parameter $t$ can be defined for WKB wavefunctions so that the equation for the perturbations $\psi_n$ can be written as a Schr$\ddot{\rm o}$dinger equation. 
If $S$ is the action for the mean values $\alpha, \phi$, by defining $y \equiv \left(\partial S\slash \partial \alpha\right)\slash \left(\partial S\slash \partial \phi\right)\sim \dot{\alpha}\slash \dot{\phi}$, we can write\cite{halliwellhawking,kiefer}:
\bea
\label{eq:pertschr}
\psi_n &=& e^{\frac{\alpha}{2}} \exp\left(i \frac{3}{2 y} \frac{\partial S}{\partial \phi} f_n^2\right)\psi_n^{(0)}\nonumber \\
i\frac{\partial \psi_n^{(0)}}{\partial t}  &=& e^{-3 \alpha} \left\{-\frac{1}{2} \frac{\partial^2}{\partial f_n^2} + U(\alpha,\phi) f_n^2\right\} \psi_n^{(0)}\nonumber \\
U(\alpha,\phi) &=&e^{6\alpha} \left\{(\frac{n^2-1}{2})e^{-2\alpha} +\frac{m^2}{2} +\right .\nonumber \\
&+& \left . 9m^2 y^{-2}\phi^2
-6m^2 y^{-1}\phi\right\}.
\eea

During inflation, $S\approx-1\slash 3\ m e^{3\alpha} \phi_{\rm inf}$, with $\phi_{\rm inf}$ the value of the field during inflation, we have $y=3\phi_{\rm inf}$ and $U=U_{-}$. Long wavelength matter fluctuations are amplified during inflation and driven away from their ground state. After inflation, when the wavepacket is in an oscillatory regime, $y$ is large and the potential $U(\alpha, \phi)$ changes from $U_{-}(\alpha, \phi)$ to $U_{+}(\alpha, \phi)$, where 
\be
\label{eq:posnegU}
U_{\pm} (\alpha, \phi) \sim e^{6\alpha} \left[\frac{n^2-1}{2}e^{-2\alpha}\pm \frac{m^2}{2}\right].
\ee

as can be seen from Eq.\ref{eq:pertschr}.
From Eq.(\ref{eq:pertschr}) we see that during inflation, the patches that have $U(\alpha, \phi)<0$, which can happen for small enough physical wave vector $k_n = n e^{-\alpha}$, develop tachyonic instabilities due to the growth of perturbations: $\psi_n \simeq e^{-\mu_n \alpha} e^{i \mu_n \phi}$, where $-\mu_n^2 = U(\alpha,\phi)f_n^2$. These trajectories in phase space {\em cannot } give rise to an inflationary universe, since they are damped in the intrinsic time $\alpha$ and so such modes do {\em not} contribute to the phase space of inflationary initial conditions. 

The damping of these wavefunctions on phase space variables is correlated with the tachyonic, Jeans-like instabilities of the corresponding mode $f_n$ in real spacetime; when $U(\alpha, \phi)<0$, $f_n\sim e^{\pm \mu_n t}$, while for $U(\alpha, \phi)>0$, the long wavelength matter perturbations $f_n$ are frozen in. By the equation of motion for $\phi,f_n$, obtained by varying the action with respect to these variables, we can see that the massive perturbation $f_n$ in real spacetime are developing an instability,i.e growing with time in the tachyonic case $U<0$ universes,

\begin{equation}
\ddot{f}_n + 3H\dot{f}_n + \frac{U_{\pm}}{a_I^{3}} f_n =0,
\label{eq:modegrowth}
\end{equation}
where the inflation scale factor is $a_I =e^{3\alpha_I}$ and $U_{\pm}$ denotes the potential/(tachyonic) mass term case, Eqn.\ref{eq:posnegU}.

For $U<0$, one obtains growing and decaying solution from Eq.\ref{eq:modegrowth} in spacetime, that go roughly as $f_n \simeq e^{\pm \mu t}$. When $U>0$ then the $f_n$ are nearly frozen as in the standard perturbation theory case for superHubble wavelength modes.

This shows that, for damped universe solution in configuration space,with $U<0, \Psi\simeq e^{-\mu\alpha}$, the perturbation modes in real space $f_n$ have a fast rate of growth. This corresponds to a universe that is collapsing instead of inflating due to the backreaction of massive superHubble perturbations $f_n$ which are coupled to the 3-geometry gravitationally via $U(\alpha,\phi)$. Note that the superHubble modes are not adiabatic and they do not re-enter in their ground state but rather in a highly excited state, thus they have a significant contribution.

The dynamics of the initial patches, considered here to contain both matter and gravitational degrees of freedom, leads us to the following {\it superselection rule}: all initial inflationary patches, characterized by values of the scale factor $a_{\rm inf}$ and Hubble parameter $h_{\rm inf}\equiv \sqrt{2\slash 3\pi} H_{\rm inf}\slash M_{\rm Planck} $ for which $U<0$ will collapse due to the backreaction of the superhorizon modes satisfying $k_n\leq m$. Since the backreaction effects due to modes with wavenumber $n$ scale as $a^{-2}$, patches for which $U>0$ will start to inflate and the backreaction effects will be inflated away. The {\it surviving patches} are then exactly those with 
\be
\label{eq:patchinf}
m^2 \phi_{\rm inf}^2 \simeq h_{\rm inf}^2 \geq k_n^2 = \left(\frac{n}{a}\right)^2\geq m^2\Rightarrow \phi_{\rm inf}\geq 1.
\ee
Eq.\ref{eq:patchinf} is exactly the superselection rule we wished to derive from the quantum dynamics of gravity, namely: the quantum dynamics of the backreacting modes wipes out from the phase space all those initial patches that cannot support inflation! This compression of the phase space of inflationary initial conditions implies that gravitational dynamics does {\em not} conserve the volume of the phase space, i.e. Liouville's theorem does not hold so that $\left[\hat{H}, \rho_{\rm red}\right]\neq 0$, (see \cite{richlaura2}). 

The entropy can be obtained by taking the log of the action above as is usual. However to simplify a rather messy expression for the action in our Master equation, let us take  the limit when the massive modes $f_n$ collapse into one black hole. Then we can write an  approximate expression for the entropy $S$ of the system
\be
S \simeq (r_I -r_{f_{n}})^2,\  r_I \simeq H_I^{-1},\  r_{f_n} \simeq H_I^{-3/2}\langle \phi_I\sqrt{U}\rangle.
\label{desitterblackhole}
\ee
where up to numerical factors of order 1, $r_I$ denotes the De-Sitter horizon of the inflationary patches with Hubble parameter $H_I$ and $r_{f_N}$ the horizon of the ``black hole'' made up from the $f_n$ modes.
\section{Remarks}
\label{sec:conc}

 What we have learned in this program by incorporating the quantum dynamics of matter and gravitational degrees of freedom is that : independent of the model, since a generic universe will always contain both, matter and gravitational degrees of freedom, then their opposing tendencies towards equlibrium drive the system out-of-equilibrium and the corresponding phase space to non-ergodic behaviour; generically, this non-equilibrium dynamics gives rise to a {\it superselection rule for the Initial Conditions}, as shown here, since non-ergodicity compresses the available volume $V$ of phase space to a subclass of survivor universes only; the low entropy, which is a function of the phase space volume $S\simeq log[V]$, results from the reduction of phase space, thus provides an explanation for the observed arrow of time and the superselection Eq.\ref{eq:patchinf}, of only high energy patches that can survive the gravitational instability of collapsing massive modes $f_n$.
 
Despite having made use of a particular model of the landscape to arrive at our results we would argue that our results should have wider applicability. The landscape minisuperspace serves mostly to provide a concrete realization of our approach, specifically the scales $M_*, M_{\rm SUSY}$ for the widths of the initial wavepackets. The rest of the quantum cosmological calculation based on backreaction and the Master Equation is general and could be applied to any phase space for the initial conditions once its structure was known from the underlying theory. Although aware of the problems and subtleties related to quantum cosmology, this formalism provides a calculational framework for exploring the cosmological implications of the string theory landscape, thus it provides a useful tool for making progress at present. 

How can we intrepret the emergence of the superselection rule for the initial conditions of inflation? We can try to understand and interpret these results on the basis that gravitational degrees of freedom comprise a 'negative heat capacity system' which reaches equilbrium by tending to infinity, while matter as a 'positive heat capacity' system reaches equilibrium by collapse. The combined system of matter and gravity, that describes any realistic cosmology, thus can not be in thermal equlibrium. The investigation of these systems must be carried out by out-of equilibrium methods. If inflation starts at high enough energies to overcome the collapsing tendency from the backreaction of matter degrees of freedom then we have a 'survivor' universe, Eq.\ref{eq:patchinf}, that continues to expand despite the backreaction term. The collapsing initial patches that can't overcome the matter backreaction can not give rise to expanding spacetimes thus they become irrelevant points on the phase space. As a result, the phase space volume decreases to only that small corner of its original volume which contains 'survivor' patches only, namely the very low entropy initial conditions. The initial entropy, roughly $S\simeq log[V]$, with $V$ the volume of phase space, becomes very small since the non-ergodic phase space has effectively reduced its original volume, due to the out-of equlibrium dynamics as shown here. 

As these results indicate, taking into account the quantum dynamics of gravity may provide the first step to reconciling the friction between inflationary initial conditions, entropy and arrow of time, as well as to shed light in the validity of some of our assumptions such as ergodicity and causal patch physics \cite{richlaura2}.

Can these ideas be tested? As mentioned earlier, an initial mixed state can not evolve into a pure state under unitary evolution. Therefore it is entirely possible that traces of quantum entaglement may have survived and left their imprint on astrophysical observables, an issue currently under investigation. Weak lensing surveys of large scale structure may provide a wealth of new information for cosmology that will complement the exquisite measurements of CMB anisotropies\cite{wmap} .

\begin{acknowledgments}
I would like to thank the organizers of DSU 2006 -Madrid for their kind hospitality.
LMH was suported in part by DOE grant DE-FG02-06ER41418 and NSF grant PHY-0553312.
\end{acknowledgments}


\end{document}